\documentclass[letter]{aa} 

\usepackage{graphicx}
\usepackage{txfonts}
\usepackage[colorlinks=false]{hyperref}
\begin{document}

   \title{Azimuthal brightness modulation reveals hidden rings in CI Tau}

   \author{C. E. Scardoni
          \inst{1}
          \and 
          G. P. Rosotti
          \inst{1}
          \and
          F. Zagaria
          \inst{2}
          \and
          A. Ribas
          \inst{3}
          \and
          M. Villenave
          \inst{4}
          \and
          C. J. Clarke
          \inst{5}
          \and
          R. A. Booth
          \inst{6}
          }

   \institute{Dipartimento di Fisica, Università degli Studi di Milano, Via Celoria 16, Milano, 20133, Italy\\
              \email{chiara.scardoni@unimi.it}
         \and
             Max-Planck Institute for Astronomy (MPIA), Königstuhl 17, 69117 Heidelberg, Germany.
         \and
             Astronomy Unit, Department of Physics and Astronomy, Queen Mary University of London, Mile End Road, London E1 4NS, UK
         \and
             Univ. Grenoble Alpes, CNRS, IPAG, F-38000 Grenoble, France
         \and
             Institute of Astronomy, University of Cambridge, Madingley Road, Cambridge, CB3 0HA, UK
        \and School of Physics and Astronomy, University of Leeds, Leeds, LS2 9JT, UK
             }

   \date{Received XXX; accepted XXX}

\abstract
{Protoplanetary disks often host substructures such as rings and gaps, which trace key processes in planet formation and dust evolution. However, narrow rings may remain unresolved due to limited observational resolution, hiding critical information about early planetesimal formation and the amount of dust present.}
{{We apply the azimuthal brightness modulation method, based on the modulation produced by multiple unresolved optically thick rings embedded in an optically thin background \citep{Scardoni+2024}, to multi-wavelength ALMA observations of CI Tau to identify unresolved rings and constrain their geometry and optical depth.}}
{We analysed CI Tau archival ALMA continuum observations in bands 3, 6, and 7, extracting azimuthal brightness profiles along narrow annuli and comparing them with forward modeled synthetic observations of inclined disks containing unresolved rings.}
{We detect the azimuthal signature at $\sim22$ au in all three bands, consistent with unresolved, optically thick rings embedded in a lower optical depth background. Multi wavelength modelling constrains the rings' geometry and optical depth, consistent with conditions expected for streaming instability and early planetesimal formation.}
{Our results demonstrate the applicability of this azimuthal signature technique to real disks, reveal fine scale dust substructures in CI Tau, and illustrate a new method to study the early stages of planet formation below the nominal resolution limit.}


   \maketitle
%

\section{Introduction}
High resolution ALMA observations reveal prominent substructures in protoplanetary disks \citep[e.g.][]{ALMA2015,Andrews2018,Andrews2020,Bae+2023,Drazkowska+2023, Long+2018, Huang+2018}, yet even narrower dust structures may remain unresolved. As angular resolution improves, narrower rings can emerge \citep[e.g.][]{Isella+2018, Facchini+2020}, and ALMA data may still hide narrow rings.

Unresolved dust substructures are particularly relevant as they may trace dust trapping, grain growth, and even the onset of streaming instability and planetesimal formation \citep[e.g.,][]{Youdin&Goodman2005, Johansen+2007, Carrera+2015, Yang+2017,Carrera+2021,Carrera+2022}. Dust evolution models predict that, in smooth disks, radial drift rapidly removes mm–cm dust grains from the disc \citep{Adachi+1976,Weidenschilling1977,Takeuchi+2002,Brauer+2008,Pinte&Laibe2014}, posing a major barrier to planetesimal formation. Localised pressure maxima can concentrate solids and halt radial drift \citep[e.g.][]{Pinilla+2012, Flock+2015, Riols&Lesur2018}. If smaller than the observation resolution, these dust traps may remain unresolved yet still host early planetesimal growth. {Observed DSHARP ring surface brightnesses are consistent with optically thin 1.3 mm emission, with optical depths distributed in a surprisingly narrow range just below unity \citep[e.g.,][]{Huang+2018,Dullemond+2018}.} Proposed explanations for such optical depth include dust scattering \citep{Zhu+2019}, planetesimal formation \citep{Dullemond+2018,Stammler2019}, or unresolved optically thick substructures \citep{Jennings2022a}.

Recent works focused on indirect methods and observational contraints to detect unresolved substructures and streaming instability \citep[e.g.,][]{Zagaria+2023,Scardoni+2021,Scardoni+2024}. \citet{Scardoni+2024} demonstrated that even when rings are not spatially resolved, their optical depth and viewing geometry imprint a characteristic signature: two brightness peaks at the disk minor axis. This azimuthal pattern arises purely from projection and radiative transfer effects, making it a powerful diagnostic to identify unresolved dust rings, as other mechanisms produce different signatures; e.g., optically thin rings produce bright emission along the major axis \citep{Doi&Kataoka2021}, {while optically thick cavities in inclined discs produce one-sided brightness maxima along the minor axis \citep{Ribas+2024}.}

In this Letter, we report the first observational detection of this azimuthal brightness modulation in the CI Tau disk. Our results show that narrow, optically thick unresolved rings can exist below the nominal ALMA resolution and leave a measurable imprint. This demonstrates that unresolved sub-rings remain a viable explanation for part of the DSHARP optical depth puzzle, suggesting that fine-scale dust concentrations may be more common than previously inferred from imaging alone. These narrow, dense ring-like structures are potentially linked to dust trapping or streaming instability, providing a new observational pathway to probe the earliest stages of planetesimal formation.

\section{Data}
\label{sec:Data}
\begin{figure}
   \centering
   \includegraphics[width=0.8\linewidth]{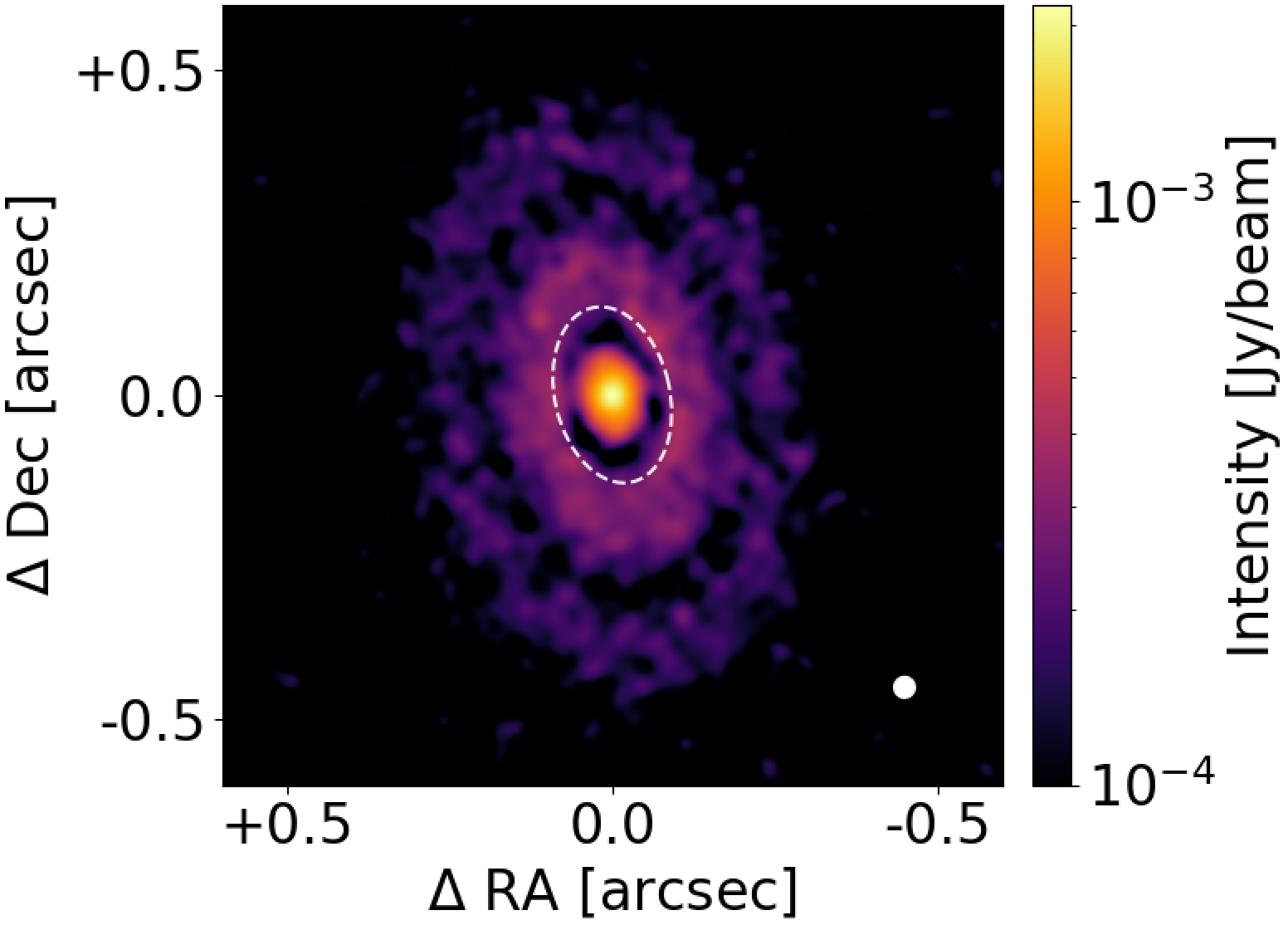}%
   \caption{{CI Tau in ALMA in Band 6. The white circle indicates the beam; the dashed ellipse shows the location where the signature is detected.}}
   \label{Fig:CITau}
\end{figure}

\begin{figure*}
   \centering
   \includegraphics[width=0.85\linewidth]{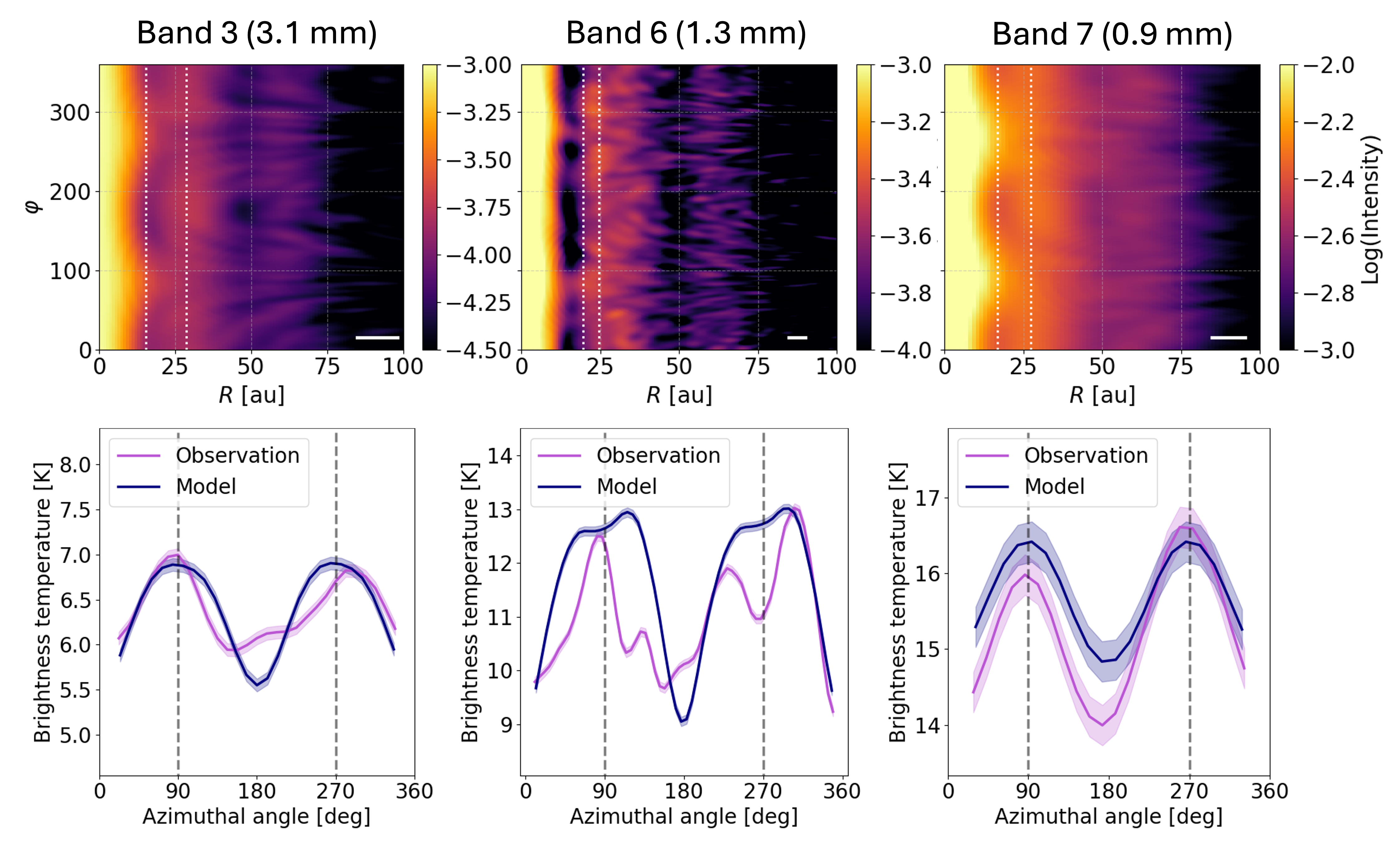}%
   \caption{{Upper row: 2D intensity maps for CI Tau at ALMA bands 3, 6, 7. The white bar in each panel shows the beam size. White dotted lines mark the profile radial averaging intervals: $22.0\pm 5.3$ (Band 3), $22.0\pm 2.6$ (Band 6), $22.0\pm 6.6$ (Band 7). Lower row: CI Tau azimuthal brightness temperature profiles in ALMA Band 3 (left), 6 (centre), 7 (right) shown in magenta, with the corresponding model profiles overplotted in blue. {The blue curves are illustrative examples from the model grid that qualitatively reproduce the observed signature (not obtained from a formal best-fit procedure).} Shaded areas show profile uncertainties. Grey dashed lines mark the minor axis.}}   
   \label{Fig:AzimuthalProfiles}
\end{figure*}

CI Tau is a $\sim$2 Myr \citep{Gangi+2022}, Sun-like T Tauri star at 160 pc \citep{GaiaCollaborationVallenari2023}. Its inclined ($i \sim 50^{\circ}$, \citealt{Clarke+2018}) disk shows resolved rings and gaps \citep{Konishi+2018, Clarke+2018, Zagaria+2025, Long+2018, Rosotti+2021}.
We analysed CI Tau archival ALMA continuum observations in bands 3 (3.1 mm, \citealt{Zagaria+2025}), 6 (1.3 mm, \citealt{Konishi+2018,Clarke+2018}), and 7 (0.9 mm, \citealt{Rosotti+2021}). Data were averaged in 20 s bins and imaged following the CLEAN algorithm \citep{Hogbom1974} in CASA \texttt{tclean} \citep{CasaTeam2022}, using an elliptical mask centreed on the source ($1\farcs5 \times 0\farcs9$, PA $11^{\circ}$, \citealt{Clarke+2018}). We adopted a multiscale deconvolver for Band 6-7, and multiscale multifrequency synthesis for Band 3 (as the data cover a significant fraction of the average observing frequency) with scales corresponding to a point source and multiples of the beam. Our cellsize is 1/8 of the beam semi-minor axis and our image size is 2400 pixels. We used a conservative loop gain of 0.02 and a threshold of 1$\sigma$. We used a Briggs weighting scheme \citep{Briggs1995} and a combination of $uv$-taper and robust parameters to achieve the smallest possible nearly circular beam. Images were smoothed to our circular beam with the task \texttt{imsmooth}. Circular beams ensure that any azimuthal intensity signature is not due to beam geometry. We obtained synthesised beams of 66, 32, 83 mas (Band 3, 6, 7) and RMS noise of $7.73\times10^{-3}$, $3.41\times10^{-2}$, $1.08\times10^{-1}$ mJy beam$^{-1}$  (Band 3, 6, 7) measured over an emission free annular region between $4\farcs0$ and $6\farcs0$ around our target.

\section{Detection of the azimuthal signature}
\label{sec:Detection of the azimuthal signature}
To search for the azimuthal signature,\footnote{{The `signature' is a double peak in the azimuthal brightness profile at the minor axis, above the local noise and robust to radial binning.}} we deprojected the disk {using inclination $49.24^{\circ}$ and position angle  $11.28^{\circ}$}(\citealt{Clarke+2018}). We then radially averaged the emission in concentric annuli at all disk radii to extract the azimuthal brightness temperature profiles; the radial width of each annulus was set to the size of the circular beam (5.3 au in band 3, 2.6 au in band 6, 6.6 au in band 7). A five-point moving average was applied azimuthally.

{We systematically scanned all radial profiles, as the signature, being associated to unresolved structures, could be present at any radius.} {We found the azimuthal signature at 22 au (magenta lines in \figureautorefname~\ref{Fig:AzimuthalProfiles}).}\footnote{{We do not find a similar signature at other radii.}} This radius is close to a surface brightness maximum just outside a deep gap seen in high-resolution Band 6 imaging \citep{Clarke+2018}.
The upper panels show the corresponding polar intensity maps; white dashed lines mark the radial regions used in the profiles.
All three bands exhibit a clear double peak profile, with maxima at $\sim 90^\circ$ and $\sim 270^\circ$.\footnote{{$0^\circ$-$180^\circ$ is the major axis; $90^\circ$ ($270^\circ$) is the near (far) side minor axis.}} This is the expected signature of an unresolved, optically thick substructures embedded in an optically thin background, as geometrical projection increases the apparent area (and thus brightness) of optically thick emission at the minor axis \citep{Scardoni+2024}.
In band 3, the minor axis peak reaches $T_{\rm B3}^{\rm max} \sim 7~{\rm K}$, with $\Delta T_{\rm B3} \sim 1~{\rm K}$ above the major axis. Band 6 has a stronger signal: $T_{\rm B6}^{\rm max} \sim 13~{\rm K}$ and $\Delta T_{\rm B6} \sim 4~{\rm K}$. In band 7, we find $\Delta T_{\rm B7} \sim 2.5~{\rm K}$ and $T_{\rm B7}^{\rm max} \sim 16.5~{\rm K}$.
These values serve as inputs for our forward-modelling (\sectionautorefname~\ref{sec:modelling the azimuthal signature}) but should not be taken as exact, as the rings are unresolved and amplitudes are affected by beam dilution.
The presence of the double peak at the same radius ($\sim 22$~au) across all three bands strongly supports the interpretation of unresolved, optically thick rings.

\section{Modelling the azimuthal signature}
\label{sec:modelling the azimuthal signature}
To study the signature, we used a forward modelling based on synthetic observations. This enables direct comparison between unresolved ring models and the observations, accounting for beam convolution, projection, and interferometric imaging.
\begin{figure}
   \centering
   \includegraphics[width=0.73\linewidth]{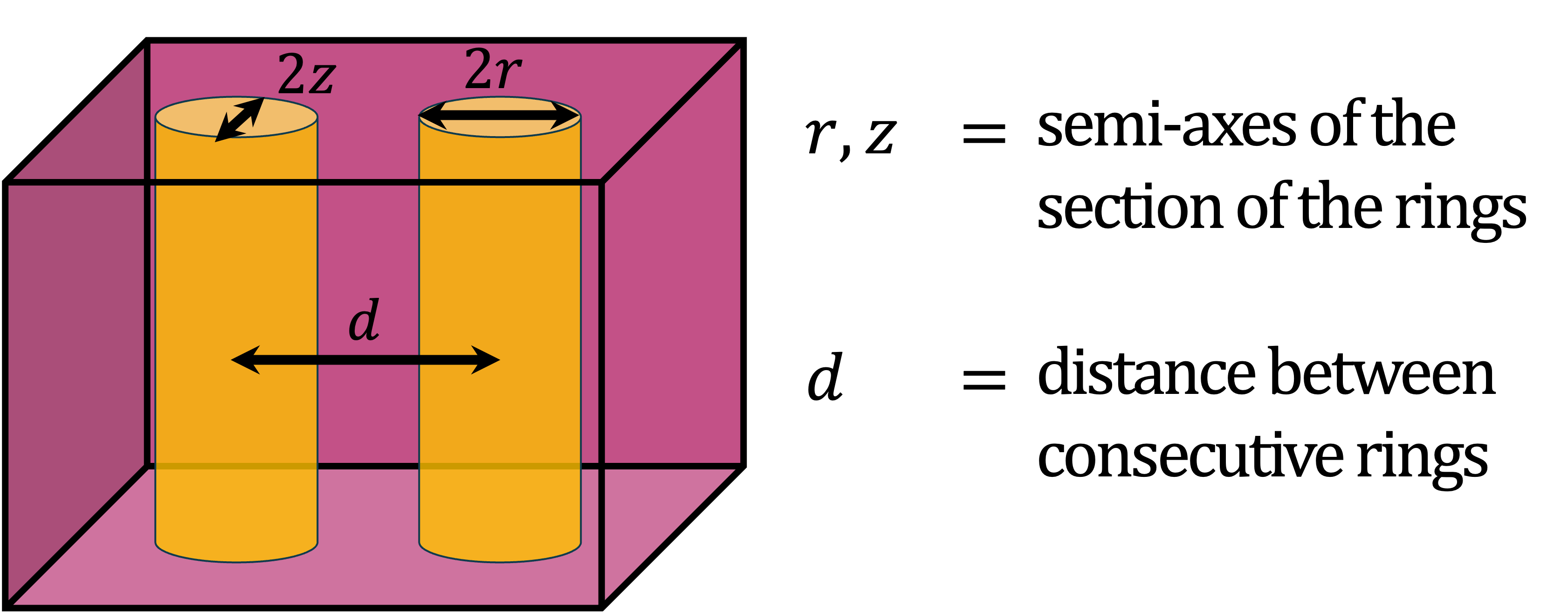}%
   \caption{{Sketch of the geometry of a section of the unresolved rings.}}
   \label{Fig:geomscheme}
\end{figure}

We used a disk profile $\Sigma(R) = 2200\ (R/{\rm au})^{-0.5}\,$g\,cm$^{-2}$, introducing a series of narrow rings at $22$ au and a depleted region between 10$-$19 au. {The disc was inclined by $50^\circ$, and its emission computed assuming midplane temperature $T(R) = 120\ (R/{\rm au})^{-3/7}$ K \citep{Chiang&Youdin2010}.} We focused on the key parameters governing the signature {(\figureautorefname~\ref{Fig:geomscheme})}:
(i) the background optical depth $\tau_{\rm bg}$; (ii) the surface coverage of the optically thick component, via the ring diameter-to-spacing ratio $2r/d$; (iii) the ring aspect ratio $r/z$, setting the projected optically thick area. {The rings' optical depth was fixed at $\tau_{\rm ring}=10$ in Band 6-7, where they are expected to be optically thick; in Band 3 we explored lower $\tau_{\rm ring}$, expecting optically thinner emission (see \appendixautorefname~\ref{appendix:Wavelength dependence of the azimuthal signature})}. Uncertainties on brightness temperature were estimated from the image RMS converted to brightness temperature and scaled by the square root of the number of independent synthesised beams within the averaging annulus.

{We performed a grid search on band 6 data, varying $\tau_{\rm bg}$, $2r/d$, and $r/z$} to identify combinations reproducing the minor axis double peak. For each model, we computed the analytical brightness distribution and generated synthetic observations with SIMIO \citep{Kurtovic2024} matching the $uv$-coverage of the data and the circular beams defined in \sectionautorefname~\ref{sec:Data}. Azimuthal profiles were extracted as for the observations and compared to the data.
Blue lines in the central lower panel of \figureautorefname~\ref{Fig:AzimuthalProfiles} show an illustrative model ($\tau_{\rm bg}=0.01$, $2r/d=0.1$, $r/z=0.17$), matching the brightness temperature profiles.
Several models provide similarly good fits (\appendixautorefname~\ref{appendix:geometrical_parameters}), reflecting the degeneracy of a three parameter model ($\tau_{\rm bg}$, $2r/d$, $r/z$) constrained by two observables ($T_{\rm minor}$, $\Delta T$). Yet, the modelling robustly confirms that the observed profile is consistent with emission from narrow, optically thick rings in an optically thin background.

{Once constrained the disk geometry from Band 6, we focused on Bands 3 and 7, varying only $\tau_{\rm bg}$ and $\tau_{\rm ring}$. The weaker signature observed in Band 3 requires marginally optically thick rings ($\tau_{\rm ring} \lesssim 5$) embedded in a very optically thin background ($\tau_{\rm bg} \sim 10^{-3}$), consistent with expectations at longer wavelengths. 
Conversely, in band 7, the background must be moderately optically thin ($\tau_{\rm bg} \sim$ 0.1), while the rings remain optically thick.}

The CI Tau azimuthal brightness pattern thus indirectly probes unresolved, optically thick dust structures. The strength of the signature varies with wavelength, reflecting the contrast between the rings and the surrounding disk \citep{Scardoni+2024}: it peaks when the rings are optically thick and the background is optically thin (Band 6), weakens at longer wavelengths as the rings approach marginal optical thickness (Band 3), and is reduced at shorter wavelengths as the background becomes more opaque (Band 7). Consistently, we find $\Delta T_{\rm B3} < \Delta T_{\rm B7} < \Delta T_{\rm B6}$. {Two order of magnitude difference are needed in $\tau_{\rm bg}$ from Band 3 to Band 6, corresponding to opacity index $\beta\sim 2$ ($k_{\nu}\propto \nu^{\beta}$), consistent with a background made of small grains.} While some degeneracy remains due to the limited number of observables (\appendixautorefname~\ref{appendix:geometrical_parameters}), our results show that unresolved, optically thick rings can explain the azimuthal asymmetry in CI Tau.

\section{Discussion}
\label{sec:discussion and implications}
\subsection{Origin of CI Tau’s unresolved ring-like dust structures}
\textbf{Streaming instability (SI):} SI naturally produces dense, radially thin, azimuthally elongated filaments made of relatively large dust grains, while smaller ones remain in the low density, optically thin background \citep[e.g.][]{Youdin&Goodman2005,Johansen+2007,Bai&Stone2010,Yang+2017}. This configuration is consistent with the narrow, optically thick rings embedded in a thinner background inferred at 22 au in CI Tau. Detecting such structures requires the SI to be in its filamentary phase, before filaments collapse into clumps. Global simulations indicate that filaments can persist for hundreds to thousands of orbits {($\sim 10^4-10^5$ yr at 22 au)} and drift inward slowly \citep{Ostertag&Flock2025}, increasing the probability of observational detection. \citet{Schafer+2024} recently confirmed the azimuthal elongation of SI filaments in global simulations, but also showed that their length is limited; SI substructures thus appear as “discontinuous rings”. Given their geometry, such features may imprint a double peaked azimuthal signature similar to that predicted by \citet{Scardoni+2024}, though noisier. Further work is required to confirm this expectation.

\textbf{Dust traps:} Localised pressure maxima, arising from planets, zonal flows, or dead-zone edges, can halt radial drift and concentrate dust \citep[e.g.,][]{Pinilla+2012,Flock+2015,Dong+2018}. The peak location just outside a gap supports this scenario, as gap edges are favourable sites for dust accumulation. {If the dust trap(s) create multiple unresolved rings, they may cause the 22 au feature} and may in turn provide the conditions for SI or other planetesimal formation mechanisms. More speculatively, a previously more massive ring has already undergone gravitational collapse, and the resulting planet(esimal) carved a dip at the ring centre, producing a pair of closely spaced rings.

\textbf{Other mechanisms:} Secular gravitational instability and MHD zonal flows can enhance dust locally \citep{Takahashi+2016,Bethune+2016,Riols&Lesur2018}, but typically produce structures (several au) broader than the narrow rings required by CI Tau signature. Turbulence can also produce elongated and radially narrow dust enhancements \citep{Gerosa+2023,Gerosa+2024}; the azimuthal signature generated by turbulent structures is yet characterised, and further work is needed to assess whether turbulence alone can produce the observed pattern.
{Simulations also suggest that WInDI (Warp-Induced Dust Instability), a dust instability in warped disks driven by oscillatory warp-induced gas motions,} can produce narrow dust overdensities on radial scales of a few to several AU, manifesting as broken ring‑like features in the dust distribution \citep{Aly+2024}; these structures may produce azimuthal signatures similar to that in CI Tau, but require the presence of a warp.
Purely geometrical effects (e.g., eccentricity or warps) can also generate double peaked azimuthal emission; {however, aligning with the minor axis requires fine-tuned geometry and predicts a wavelength independent signature (contrary to observations), making this scenario less likely than SI (see \appendixautorefname~\ref{appendix:warped discs}).}

\subsection{Implications and prospects}
The azimuthal signature at 22 au is consistent with the presence of narrow, optically thick rings. If interpreted as rings, their high optical depth and compact width imply efficient local dust concentration able to slow radial drift and promoting solid accumulation. Such conditions are consistent with streaming instability or other mechanisms (e.g., dust traps); the 22 au region is thus a potential site for the early stages of planetesimal formation.

This detection also suggests that similarly compact substructures may be hidden below ALMA resolution; indeed, regardless of beam size, this method can reveal sub-beam structures offering a powerful way to reveal them. Applying \citet{Scardoni+2024} method to other disks could uncover a population of sub-beam rings and guide targeted high resolution, multi wavelength follow-up observations. Whether the hidden rings are resolved in future high resolution, the signature itself constrains the geometry, dust properties, and physical scales of early dust concentration. {Applying this method to a wider set of high-resolution observations may also help constrain the lifetime of SI-induced filaments based on their detection frequency.}

\section{Conclusions}
{We report the first detection of the azimuthal brightness modulation (with emission peaks at the minor axis of an inclined disk) predicted by \citet{Scardoni+2024} in ALMA Bands 3, 6, 7 CI Tau observations. The signature, detected at 22 au, is consistent with narrow, optically thick dust rings in a low optical depth background. A single disk geometry, combined with wavelength dependent dust optical depth, reproduces the observed profiles and constrains the rings’ geometry and optical properties.}
While alternative mechanisms (e.g., warped disks, MHD perturbations) can generate dust rings, the geometry and wavelength dependence more naturally point to SI filaments or dust traps.

This work demonstrates that the azimuthal signature is a powerful tool for revealing sub-beam dust structures hidden in continuum images. Applied to other disks, this technique could uncover a broader population of unresolved rings, guiding high-resolution ALMA follow-ups and providing new constraints on the earliest stages of planetesimal and planet formation.


\begin{acknowledgements}
{We thank the referee for the helpful comments provided to this manuscript.} C.E.S and G.P.R. acknowledge support from the European Union (ERC Starting Grant diskEvol, project No. 101039651) and from Fondazione Cariplo, grant No. 2022-2017. Views and opinion expressed are, however, those of the author(s) only and do not necessarily reflect those of the European Union or European Research Council. Neither the European Union nor the granting authority can be held responsible for them. This paper makes use of the following ALMA data: ADS/JAO.ALMA\#2015.1.01207.S, ADS/JAO.ALMA\#2016.1.01370.S, ADS/JAO.ALMA\#2017.A.00014.S, and ADS/JAO.ALMA\#2018.1.00900.S, ALMA is a partnership of ESO (representing its member states), NSF (USA), and NINS (Japan), together with NRC (Canada), NSC and ASIAA (Taiwan), and KASI (Republic of Korea), in cooperation with the Republic of Chile. The Joint ALMA Observatory is operated by ESO, AUI/NRAO, and NAOJ. A.R. has received funding from the Royal Society through a University Research Fellowship grant number URF\textbackslash R1\textbackslash 241791. CJC has been supported by the UK Science and Technology Research Council (STFC) via the consolidated grant ST/W000997/1.
R.A.B. has received funding from the Royal Society through a University Research Fellowship grant number URF\textbackslash R1\textbackslash 211799.

\end{acknowledgements}

%
\bibliographystyle{aa} 
\bibliography{references} 

@ARTICLE{Huang+2018,
       author = {{Huang}, Jane and {Andrews}, Sean M. and {Dullemond}, Cornelis P. and {Isella}, Andrea and {P{\'e}rez}, Laura M. and {Guzm{\'a}n}, Viviana V. and {{\"O}berg}, Karin I. and {Zhu}, Zhaohuan and {Zhang}, Shangjia and {Bai}, Xue-Ning and {Benisty}, Myriam and {Birnstiel}, Tilman and {Carpenter}, John M. and {Hughes}, A. Meredith and {Ricci}, Luca and {Weaver}, Erik and {Wilner}, David J.},
        title = "{The Disk Substructures at High Angular Resolution Project (DSHARP). II. Characteristics of Annular Substructures}",
      journal = {\apjl},
         year = 2018,
        month = dec,
       volume = {869},
       number = {2},
          eid = {L42},
        pages = {L42},
          doi = {10.3847/2041-8213/aaf740},
archivePrefix = {arXiv},
       eprint = {1812.04041},
 primaryClass = {astro-ph.EP},
       adsurl = {https://ui.adsabs.harvard.edu/abs/2018ApJ...869L..42H}
}

@ARTICLE{Brauer+2008,
       author = {{Brauer}, F. and {Dullemond}, C.~P. and {Henning}, Th.},
        title = "{Coagulation, fragmentation and radial motion of solid particles in protoplanetary disks}",
      journal = {\aap},
         year = 2008,
        month = mar,
       volume = {480},
       number = {3},
        pages = {859-877},
          doi = {10.1051/0004-6361:20077759},
archivePrefix = {arXiv},
       eprint = {0711.2192},
 primaryClass = {astro-ph},
       adsurl = {https://ui.adsabs.harvard.edu/abs/2008A&A...480..859B}
}

@ARTICLE{Pinte&Laibe2014,
       author = {{Pinte}, C. and {Laibe}, G.},
        title = "{Diversity in the outcome of dust radial drift in protoplanetary discs}",
      journal = {\aap},
         year = 2014,
        month = may,
       volume = {565},
          eid = {A129},
        pages = {A129},
          doi = {10.1051/0004-6361/201220545},
archivePrefix = {arXiv},
       eprint = {1403.1587},
 primaryClass = {astro-ph.SR},
       adsurl = {https://ui.adsabs.harvard.edu/abs/2014A&A...565A.129P}
}

@ARTICLE{Takeuchi+2002,
       author = {{Takeuchi}, Taku and {Lin}, D.~N.~C.},
        title = "{Radial Flow of Dust Particles in Accretion Disks}",
      journal = {\apj},
         year = 2002,
        month = dec,
       volume = {581},
       number = {2},
        pages = {1344-1355},
          doi = {10.1086/344437},
archivePrefix = {arXiv},
       eprint = {astro-ph/0208552},
 primaryClass = {astro-ph},
       adsurl = {https://ui.adsabs.harvard.edu/abs/2002ApJ...581.1344T}
}

@ARTICLE{Adachi+1976,
       author = {{Adachi}, I. and {Hayashi}, C. and {Nakazawa}, K.},
        title = "{The gas drag effect on the elliptical motion of a solid body in the primordial solar nebula.}",
      journal = {Progress of Theoretical Physics},
         year = 1976,
        month = dec,
       volume = {56},
        pages = {1756-1771},
          doi = {10.1143/PTP.56.1756},
       adsurl = {https://ui.adsabs.harvard.edu/abs/1976PThPh..56.1756A}
}

@ARTICLE{Chiang&Youdin2010,
       author = {{Chiang}, E. and {Youdin}, A.~N.},
        title = "{Forming Planetesimals in Solar and Extrasolar Nebulae}",
      journal = {Annual Review of Earth and Planetary Sciences},
         year = 2010,
        month = may,
       volume = {38},
        pages = {493-522},
          doi = {10.1146/annurev-earth-040809-152513},
archivePrefix = {arXiv},
       eprint = {0909.2652},
 primaryClass = {astro-ph.EP},
       adsurl = {https://ui.adsabs.harvard.edu/abs/2010AREPS..38..493C}
}

@ARTICLE{Stammler2019,
       author = {{Stammler}, Sebastian M. and {Dr{\k{a}}{\.z}kowska}, Joanna and
         {Birnstiel}, Til and {Klahr}, Hubert and {Dullemond}, Cornelis P. and
         {Andrews}, Sean M.},
        title = "{The DSHARP Rings: Evidence of Ongoing Planetesimal Formation?}",
      journal = {\apjl},
         year = 2019,
        month = oct,
       volume = {884},
       number = {1},
          eid = {L5},
        pages = {L5},
          doi = {10.3847/2041-8213/ab4423},
archivePrefix = {arXiv},
       eprint = {1909.04674},
 primaryClass = {astro-ph.EP},
       adsurl = {https://ui.adsabs.harvard.edu/abs/2019ApJ...884L...5S}
}

@ARTICLE{Andrews2018,
       author = {{Andrews}, Sean M. and {Huang}, Jane and {P{\'e}rez}, Laura M. and
         {Isella}, Andrea and {Dullemond}, Cornelis P. and
         {Kurtovic}, Nicol{\'a}s T. and {Guzm{\'a}n}, Viviana V. and
         {Carpenter}, John M. and {Wilner}, David J. and {Zhang}, Shangjia and
         {Zhu}, Zhaohuan and {Birnstiel}, Tilman and {Bai}, Xue-Ning and
         {Benisty}, Myriam and {Hughes}, A. Meredith and {{\"O}berg}, Karin I. and
         {Ricci}, Luca},
        title = "{The Disk Substructures at High Angular Resolution Project (DSHARP). I. Motivation, Sample, Calibration, and Overview}",
      journal = {\apjl},
         year = 2018,
        month = dec,
       volume = {869},
       number = {2},
          eid = {L41},
        pages = {L41},
          doi = {10.3847/2041-8213/aaf741},
archivePrefix = {arXiv},
       eprint = {1812.04040},
 primaryClass = {astro-ph.SR},
       adsurl = {https://ui.adsabs.harvard.edu/abs/2018ApJ...869L..41A}
}

@ARTICLE{Yang+2017,
       author = {{Yang}, C. -C. and {Johansen}, A. and {Carrera}, D.},
        title = "{Concentrating small particles in protoplanetary disks through the streaming instability}",
      journal = {\aap},
         year = 2017,
        month = oct,
       volume = {606},
          eid = {A80},
        pages = {A80},
          doi = {10.1051/0004-6361/201630106},
archivePrefix = {arXiv},
       eprint = {1611.07014},
 primaryClass = {astro-ph.EP},
       adsurl = {https://ui.adsabs.harvard.edu/abs/2017A&A...606A..80Y}
}

@ARTICLE{Carrera+2015,
       author = {{Carrera}, Daniel and {Johansen}, Anders and {Davies}, Melvyn B.},
        title = "{How to form planetesimals from mm-sized chondrules and chondrule aggregates}",
      journal = {\aap},
         year = 2015,
        month = jul,
       volume = {579},
          eid = {A43},
        pages = {A43},
          doi = {10.1051/0004-6361/201425120},
archivePrefix = {arXiv},
       eprint = {1501.05314},
 primaryClass = {astro-ph.EP},
       adsurl = {https://ui.adsabs.harvard.edu/abs/2015A&A...579A..43C}
}

@ARTICLE{Weidenschilling1977,
       author = {{Weidenschilling}, S.~J.},
        title = "{Aerodynamics of solid bodies in the solar nebula.}",
      journal = {\mnras},
         year = 1977,
        month = jul,
       volume = {180},
        pages = {57-70},
          doi = {10.1093/mnras/180.1.57},
       adsurl = {https://ui.adsabs.harvard.edu/abs/1977MNRAS.180...57W}
}

@ARTICLE{Youdin&Goodman2005,
       author = {{Youdin}, Andrew N. and {Goodman}, Jeremy},
        title = "{Streaming Instabilities in Protoplanetary Disks}",
      journal = {\apj},
         year = "2005",
        month = "Feb",
       volume = {620},
       number = {1},
        pages = {459-469},
          doi = {10.1086/426895},
archivePrefix = {arXiv},
       eprint = {astro-ph/0409263},
 primaryClass = {astro-ph},
       adsurl = {https://ui.adsabs.harvard.edu/abs/2005ApJ...620..459Y}
}

@ARTICLE{Johansen+2007,
       author = {{Johansen}, Anders and {Oishi}, Jeffrey S. and {Mac Low}, Mordecai-Mark and
         {Klahr}, Hubert and {Henning}, Thomas and {Youdin}, Andrew},
        title = "{Rapid planetesimal formation in turbulent circumstellar disks}",
      journal = {\nat},
         year = "2007",
        month = "Aug",
       volume = {448},
       number = {7157},
        pages = {1022-1025},
          doi = {10.1038/nature06086},
archivePrefix = {arXiv},
       eprint = {0708.3890},
 primaryClass = {astro-ph},
       adsurl = {https://ui.adsabs.harvard.edu/abs/2007Natur.448.1022J}
}

@ARTICLE{Bai&Stone2010,
       author = {{Bai}, Xue-Ning and {Stone}, James M.},
        title = "{The Effect of the Radial Pressure Gradient in Protoplanetary Disks on Planetesimal Formation}",
      journal = {\apjl},
         year = "2010",
        month = "Oct",
       volume = {722},
       number = {2},
        pages = {L220-L223},
          doi = {10.1088/2041-8205/722/2/L220},
archivePrefix = {arXiv},
       eprint = {1005.4981},
 primaryClass = {astro-ph.EP},
       adsurl = {https://ui.adsabs.harvard.edu/abs/2010ApJ...722L.220B}
}

@ARTICLE{Dullemond+2018,
       author = {{Dullemond}, Cornelis P. and {Birnstiel}, Tilman and {Huang}, Jane and
         {Kurtovic}, Nicol{\'a}s T. and {Andrews}, Sean M. and
         {Guzm{\'a}n}, Viviana V. and {P{\'e}rez}, Laura M. and
         {Isella}, Andrea and {Zhu}, Zhaohuan and {Benisty}, Myriam and
         {Wilner}, David J. and {Bai}, Xue-Ning and {Carpenter}, John M. and
         {Zhang}, Shangjia and {Ricci}, Luca},
        title = "{The Disk Substructures at High Angular Resolution Project (DSHARP). VI. Dust Trapping in Thin-ringed Protoplanetary Disks}",
      journal = {\apjl},
         year = "2018",
        month = "Dec",
       volume = {869},
       number = {2},
          eid = {L46},
        pages = {L46},
          doi = {10.3847/2041-8213/aaf742},
archivePrefix = {arXiv},
       eprint = {1812.04044},
 primaryClass = {astro-ph.EP},
       adsurl = {https://ui.adsabs.harvard.edu/abs/2018ApJ...869L..46D}
}

@ARTICLE{Scardoni+2021,
       author = {{Scardoni}, Chiara E. and {Booth}, Richard A. and {Clarke}, Cathie J.},
        title = "{The effect of the streaming instability on protoplanetary disc emission at millimetre wavelengths}",
      journal = {\mnras},
         year = 2021,
        month = jun,
       volume = {504},
       number = {1},
        pages = {1495-1510},
          doi = {10.1093/mnras/stab854},
archivePrefix = {arXiv},
       eprint = {2103.12644},
 primaryClass = {astro-ph.EP},
       adsurl = {https://ui.adsabs.harvard.edu/abs/2021MNRAS.504.1495S}
}

@article{ALMA2015,
   title={THE 2014 ALMA LONG BASELINE CAMPAIGN: FIRST RESULTS FROM HIGH ANGULAR RESOLUTION OBSERVATIONS TOWARD THE HL TAU REGION},
   volume={808},
   ISSN={2041-8213},
   url={http://dx.doi.org/10.1088/2041-8205/808/1/L3},
   DOI={10.1088/2041-8205/808/1/l3},
   number={1},
   journal={The Astrophysical Journal},
   publisher={American Astronomical Society},
   author={Brogan, C. L. and Pérez, L. M. and Hunter, T. R. and Dent, W. R. F. and Hales, A. S. and Hills, R. E. and Corder, S. and Fomalont, E. B. and Vlahakis, C. and Asaki, Y. and Barkats, D. and Hirota, A. and Hodge, J. A. and Impellizzeri, C. M. V. and Kneissl, R. and Liuzzo, E. and Lucas, R. and Marcelino, N. and Matsushita, S. and Nakanishi, K. and Phillips, N. and Richards, A. M. S. and Toledo, I. and Aladro, R. and Broguiere, D. and Cortes, J. R. and Cortes, P. C. and Espada, D. and Galarza, F. and Appadoo, D. Garcia- and Ramirez, L. Guzman- and Humphreys, E. M. and Jung, T. and Kameno, S. and Laing, R. A. and Leon, S. and Marconi, G. and Mignano, A. and Nikolic, B. and Nyman, L.-A. and Radiszcz, M. and Remijan, A. and Rodón, J. A. and Sawada, T. and Takahashi, S. and Tilanus, R. P. J. and Vilaro, B. Vila and Watson, L. C. and Wiklind, T. and Akiyama, E. and Chapillon, E. and Monsalvo, I. de Gregorio- and Francesco, J. Di and Gueth, F. and Kawamura, A. and Lee, C.-F. and Luong, Q. Nguyen and Mangum, J. and Pietu, V. and Sanhueza, P. and Saigo, K. and Takakuwa, S. and Ubach, C. and Kempen, T. van and Wootten, A. and Carrizo, A. Castro- and Francke, H. and Gallardo, J. and Garcia, J. and Gonzalez, S. and Hill, T. and Kaminski, T. and Kurono, Y. and Liu, H.-Y. and Lopez, C. and Morales, F. and Plarre, K. and Schieven, G. and Testi, L. and Videla, L. and Villard, E. and Andreani, P. and Hibbard, J. E. and Tatematsu, K.},
   year={2015},
   month=jul, pages={L3} }

@ARTICLE{Jennings2022a,
       author = {{Jennings}, Jeff and {Booth}, Richard A. and {Tazzari}, Marco and {Clarke}, Cathie J. and {Rosotti}, Giovanni P.},
        title = "{A super-resolution analysis of the DSHARP survey: substructure is common in the inner 30 au}",
      journal = {\mnras},
         year = 2022,
        month = jan,
       volume = {509},
       number = {2},
        pages = {2780-2799},
          doi = {10.1093/mnras/stab3185},
archivePrefix = {arXiv},
       eprint = {2103.02392},
 primaryClass = {astro-ph.EP},
       adsurl = {https://ui.adsabs.harvard.edu/abs/2022MNRAS.509.2780J}
}

@ARTICLE{Doi&Kataoka2021,
       author = {{Doi}, Kiyoaki and {Kataoka}, Akimasa},
        title = "{Estimate on Dust Scale Height from the ALMA Dust Continuum Image of the HD 163296 Protoplanetary Disk}",
      journal = {\apj},
         year = 2021,
        month = may,
       volume = {912},
       number = {2},
          eid = {164},
        pages = {164},
          doi = {10.3847/1538-4357/abe5a6},
archivePrefix = {arXiv},
       eprint = {2102.06209},
 primaryClass = {astro-ph.EP},
       adsurl = {https://ui.adsabs.harvard.edu/abs/2021ApJ...912..164D}
}

@INPROCEEDINGS{Drazkowska+2023,
       author = {{Drazkowska}, J. and {Bitsch}, B. and {Lambrechts}, M. and {Mulders}, G.~D. and {Harsono}, D. and {Vazan}, A. and {Liu}, B. and {Ormel}, C.~W. and {Kretke}, K. and {Morbidelli}, A.},
        title = "{Planet Formation Theory in the Era of ALMA and Kepler: from Pebbles to Exoplanets}",
    booktitle = {Protostars and Planets VII},
         year = 2023,
       editor = {{Inutsuka}, S. and {Aikawa}, Y. and {Muto}, T. and {Tomida}, K. and {Tamura}, M.},
       series = {Astronomical Society of the Pacific Conference Series},
       volume = {534},
        month = jul,
        pages = {717},
          doi = {10.48550/arXiv.2203.09759},
archivePrefix = {arXiv},
       eprint = {2203.09759},
 primaryClass = {astro-ph.EP},
       adsurl = {https://ui.adsabs.harvard.edu/abs/2023ASPC..534..717D}
}

@INPROCEEDINGS{Bae+2023,
       author = {{Bae}, J. and {Isella}, A. and {Zhu}, Z. and {Martin}, R. and {Okuzumi}, S. and {Suriano}, S.},
        title = "{Structured Distributions of Gas and Solids in Protoplanetary Disks}",
    booktitle = {Protostars and Planets VII},
         year = 2023,
       editor = {{Inutsuka}, S. and {Aikawa}, Y. and {Muto}, T. and {Tomida}, K. and {Tamura}, M.},
       series = {Astronomical Society of the Pacific Conference Series},
       volume = {534},
        month = jul,
        pages = {423},
          doi = {10.48550/arXiv.2210.13314},
archivePrefix = {arXiv},
       eprint = {2210.13314},
 primaryClass = {astro-ph.EP},
       adsurl = {https://ui.adsabs.harvard.edu/abs/2023ASPC..534..423B}
}

@ARTICLE{Facchini+2020,
       author = {{Facchini}, S. and {Benisty}, M. and {Bae}, J. and {Loomis}, R. and {Perez}, L. and {Ansdell}, M. and {Mayama}, S. and {Pinilla}, P. and {Teague}, R. and {Isella}, A. and {Mann}, A.},
        title = "{Annular substructures in the transition disks around LkCa 15 and J1610}",
      journal = {\aap},
         year = 2020,
        month = jul,
       volume = {639},
          eid = {A121},
        pages = {A121},
          doi = {10.1051/0004-6361/202038027},
archivePrefix = {arXiv},
       eprint = {2005.02712},
 primaryClass = {astro-ph.EP},
       adsurl = {https://ui.adsabs.harvard.edu/abs/2020A&A...639A.121F}
}

@ARTICLE{Hogbom1974,
       author = {{H{\"o}gbom}, J.~A.},
        title = "{Aperture Synthesis with a Non-Regular Distribution of Interferometer Baselines}",
      journal = {\aaps},
         year = 1974,
        month = jun,
       volume = {15},
        pages = {417},
       adsurl = {https://ui.adsabs.harvard.edu/abs/1974A&AS...15..417H}
}

@ARTICLE{Long+2018,
       author = {{Long}, Feng and {Pinilla}, Paola and {Herczeg}, Gregory J. and {Harsono}, Daniel and {Dipierro}, Giovanni and {Pascucci}, Ilaria and {Hendler}, Nathan and {Tazzari}, Marco and {Ragusa}, Enrico and {Salyk}, Colette and {Edwards}, Suzan and {Lodato}, Giuseppe and {van de Plas}, Gerrit and {Johnstone}, Doug and {Liu}, Yao and {Boehler}, Yann and {Cabrit}, Sylvie and {Manara}, Carlo F. and {Menard}, Francois and {Mulders}, Gijs D. and {Nisini}, Brunella and {Fischer}, William J. and {Rigliaco}, Elisabetta and {Banzatti}, Andrea and {Avenhaus}, Henning and {Gully-Santiago}, Michael},
        title = "{Gaps and Rings in an ALMA Survey of Disks in the Taurus Star-forming Region}",
      journal = {\apj},
         year = 2018,
        month = dec,
       volume = {869},
       number = {1},
          eid = {17},
        pages = {17},
          doi = {10.3847/1538-4357/aae8e1},
archivePrefix = {arXiv},
       eprint = {1810.06044},
 primaryClass = {astro-ph.SR},
       adsurl = {https://ui.adsabs.harvard.edu/abs/2018ApJ...869...17L}
}

@ARTICLE{Scardoni+2024,
       author = {{Scardoni}, Chiara E. and {Booth}, Richard A. and {Clarke}, Cathie J. and {Rosotti}, Giovanni P. and {Ribas}, {\'A}lvaro},
        title = "{Seeing the Unseen: A Method to Detect Unresolved Rings in Protoplanetary Disks}",
      journal = {\apj},
         year = 2024,
        month = aug,
       volume = {970},
       number = {2},
          eid = {109},
        pages = {109},
          doi = {10.3847/1538-4357/ad55c5},
archivePrefix = {arXiv},
       eprint = {2406.11627},
 primaryClass = {astro-ph.EP},
       adsurl = {https://ui.adsabs.harvard.edu/abs/2024ApJ...970..109S}
}

@ARTICLE{Zagaria+2025,
       author = {{Zagaria}, Francesco and {Facchini}, Stefano and {Curone}, Pietro and {Williams}, Jonathan P. and {Clarke}, Cathie J. and {Ribas}, {\'A}lvaro and {Tazzari}, Marco and {Mac{\'\i}as}, Enrique and {Booth}, Richard A. and {Rosotti}, Giovanni P. and {Testi}, Leonardo},
        title = "{Multi-frequency analysis of the ALMA and VLA high resolution continuum observations of the substructured disc around CI Tau: Preference for submillimetre-sized low-porosity amorphous carbon grains}",
      journal = {\aap},
         year = 2025,
        month = oct,
       volume = {702},
          eid = {A56},
        pages = {A56},
          doi = {10.1051/0004-6361/202452986},
archivePrefix = {arXiv},
       eprint = {2507.08797},
 primaryClass = {astro-ph.EP},
       adsurl = {https://ui.adsabs.harvard.edu/abs/2025A&A...702A..56Z}
}

@ARTICLE{Clarke+2018,
       author = {{Clarke}, C.~J. and {Tazzari}, M. and {Juhasz}, A. and {Rosotti}, G. and {Booth}, R. and {Facchini}, S. and {Ilee}, J.~D. and {Johns-Krull}, C.~M. and {Kama}, M. and {Meru}, F. and {Prato}, L.},
        title = "{High-resolution Millimeter Imaging of the CI Tau Protoplanetary Disk: A Massive Ensemble of Protoplanets from 0.1 to 100 au}",
      journal = {\apjl},
         year = 2018,
        month = oct,
       volume = {866},
       number = {1},
          eid = {L6},
        pages = {L6},
          doi = {10.3847/2041-8213/aae36b},
archivePrefix = {arXiv},
       eprint = {1809.08147},
 primaryClass = {astro-ph.EP},
       adsurl = {https://ui.adsabs.harvard.edu/abs/2018ApJ...866L...6C}
}

@ARTICLE{Konishi+2018,
       author = {{Konishi}, Mihoko and {Hashimoto}, Jun and {Hori}, Yasunori},
        title = "{Probing Signatures of a Distant Planet around the Young T-Tauri Star CI Tau Hosting a Possible Hot Jupiter}",
      journal = {\apjl},
         year = 2018,
        month = jun,
       volume = {859},
       number = {2},
          eid = {L28},
        pages = {L28},
          doi = {10.3847/2041-8213/aac6d2},
archivePrefix = {arXiv},
       eprint = {1805.07498},
 primaryClass = {astro-ph.EP},
       adsurl = {https://ui.adsabs.harvard.edu/abs/2018ApJ...859L..28K}
}

@ARTICLE{Rosotti+2021,
       author = {{Rosotti}, Giovanni P. and {Ilee}, John D. and {Facchini}, Stefano and {Tazzari}, Marco and {Booth}, Richard A. and {Clarke}, Cathie and {Kama}, Mihkel},
        title = "{High-resolution observations of molecular emission lines toward the CI Tau proto-planetary disc: planet-carved gaps or shadowing?}",
      journal = {\mnras},
         year = 2021,
        month = mar,
       volume = {501},
       number = {3},
        pages = {3427-3442},
          doi = {10.1093/mnras/staa3869},
archivePrefix = {arXiv},
       eprint = {2012.07848},
 primaryClass = {astro-ph.EP},
       adsurl = {https://ui.adsabs.harvard.edu/abs/2021MNRAS.501.3427R}
}

@ARTICLE{Ostertag&Flock2025,
       author = {{Ostertag}, Dominik and {Flock}, Mario},
        title = "{Strong clumping in global streaming instability simulations with a dusty fluid}",
      journal = {\aap},
         year = 2025,
        month = mar,
       volume = {695},
          eid = {L13},
        pages = {L13},
          doi = {10.1051/0004-6361/202453349},
archivePrefix = {arXiv},
       eprint = {2501.18424},
 primaryClass = {astro-ph.EP},
       adsurl = {https://ui.adsabs.harvard.edu/abs/2025A&A...695L..13O}
}

@ARTICLE{Isella+2018,
       author = {{Isella}, Andrea and {Huang}, Jane and {Andrews}, Sean M. and {Dullemond}, Cornelis P. and {Birnstiel}, Tilman and {Zhang}, Shangjia and {Zhu}, Zhaohuan and {Guzm{\'a}n}, Viviana V. and {P{\'e}rez}, Laura M. and {Bai}, Xue-Ning and {Benisty}, Myriam and {Carpenter}, John M. and {Ricci}, Luca and {Wilner}, David J.},
        title = "{The Disk Substructures at High Angular Resolution Project (DSHARP). IX. A High-definition Study of the HD 163296 Planet-forming Disk}",
      journal = {\apjl},
         year = 2018,
        month = dec,
       volume = {869},
       number = {2},
          eid = {L49},
        pages = {L49},
          doi = {10.3847/2041-8213/aaf747},
archivePrefix = {arXiv},
       eprint = {1812.04047},
 primaryClass = {astro-ph.SR},
       adsurl = {https://ui.adsabs.harvard.edu/abs/2018ApJ...869L..49I}
}

@ARTICLE{Pinilla+2012,
       author = {{Pinilla}, P. and {Benisty}, M. and {Birnstiel}, T.},
        title = "{Ring shaped dust accumulation in transition disks}",
      journal = {\aap},
         year = 2012,
        month = sep,
       volume = {545},
          eid = {A81},
        pages = {A81},
          doi = {10.1051/0004-6361/201219315},
archivePrefix = {arXiv},
       eprint = {1207.6485},
 primaryClass = {astro-ph.EP},
       adsurl = {https://ui.adsabs.harvard.edu/abs/2012A&A...545A..81P}
}

@ARTICLE{Flock+2015,
       author = {{Flock}, M. and {Ruge}, J.~P. and {Dzyurkevich}, N. and {Henning}, Th. and {Klahr}, H. and {Wolf}, S.},
        title = "{Gaps, rings, and non-axisymmetric structures in protoplanetary disks. From simulations to ALMA observations}",
      journal = {\aap},
         year = 2015,
        month = feb,
       volume = {574},
          eid = {A68},
        pages = {A68},
          doi = {10.1051/0004-6361/201424693},
archivePrefix = {arXiv},
       eprint = {1411.2736},
 primaryClass = {astro-ph.EP},
       adsurl = {https://ui.adsabs.harvard.edu/abs/2015A&A...574A..68F}
}

@ARTICLE{Riols&Lesur2018,
       author = {{Riols}, A. and {Lesur}, G.},
        title = "{Dust settling and rings in the outer regions of protoplanetary discs subject to ambipolar diffusion}",
      journal = {\aap},
         year = 2018,
        month = sep,
       volume = {617},
          eid = {A117},
        pages = {A117},
          doi = {10.1051/0004-6361/201833212},
archivePrefix = {arXiv},
       eprint = {1805.00458},
 primaryClass = {astro-ph.EP},
       adsurl = {https://ui.adsabs.harvard.edu/abs/2018A&A...617A.117R}
}

@ARTICLE{Kurtovic2024,
       author = {{Kurtovic}, Nicolas},
        title = "{SIMIO-continuum: Connecting simulations to ALMA observations}",
      journal = {The Journal of Open Source Software},
         year = 2024,
        month = may,
       volume = {9},
       number = {97},
          eid = {4942},
        pages = {4942},
          doi = {10.21105/joss.04942},
       adsurl = {https://ui.adsabs.harvard.edu/abs/2024JOSS....9.4942K}
}

@ARTICLE{Dong+2018,
       author = {{Dong}, Ruobing and {Li}, Shengtai and {Chiang}, Eugene and {Li}, Hui},
        title = "{Multiple Disk Gaps and Rings Generated by a Single Super-Earth. II. Spacings, Depths, and Number of Gaps, with Application to Real Systems}",
      journal = {\apj},
         year = 2018,
        month = oct,
       volume = {866},
       number = {2},
          eid = {110},
        pages = {110},
          doi = {10.3847/1538-4357/aadadd},
archivePrefix = {arXiv},
       eprint = {1808.06613},
 primaryClass = {astro-ph.EP},
       adsurl = {https://ui.adsabs.harvard.edu/abs/2018ApJ...866..110D}
}

@ARTICLE{Zagaria+2023,
       author = {{Zagaria}, F. and {Clarke}, C.~J. and {Booth}, R.~A. and {Facchini}, S. and {Rosotti}, G.~P.},
        title = "{Observing Planetesimal Formation under Streaming Instability in the Rings of HD 163296}",
      journal = {\apjl},
         year = 2023,
        month = dec,
       volume = {959},
       number = {2},
          eid = {L15},
        pages = {L15},
          doi = {10.3847/2041-8213/ad0c54},
archivePrefix = {arXiv},
       eprint = {2311.08950},
 primaryClass = {astro-ph.EP},
       adsurl = {https://ui.adsabs.harvard.edu/abs/2023ApJ...959L..15Z}
}

@ARTICLE{Takahashi+2016,
       author = {{Takahashi}, Sanemichi Z. and {Inutsuka}, Shu-ichiro},
        title = "{An Origin of Multiple Ring Structure and Hidden Planets in HL Tau: A Unified Picture by Secular Gravitational Instability}",
      journal = {\aj},
         year = 2016,
        month = dec,
       volume = {152},
       number = {6},
          eid = {184},
        pages = {184},
          doi = {10.3847/0004-6256/152/6/184},
archivePrefix = {arXiv},
       eprint = {1604.05450},
 primaryClass = {astro-ph.SR},
       adsurl = {https://ui.adsabs.harvard.edu/abs/2016AJ....152..184T}
}

@ARTICLE{Bethune+2016,
       author = {{B{\'e}thune}, William and {Lesur}, Geoffroy and {Ferreira}, Jonathan},
        title = "{Self-organisation in protoplanetary discs. Global, non-stratified Hall-MHD simulations}",
      journal = {\aap},
         year = 2016,
        month = may,
       volume = {589},
          eid = {A87},
        pages = {A87},
          doi = {10.1051/0004-6361/201527874},
archivePrefix = {arXiv},
       eprint = {1603.02475},
 primaryClass = {astro-ph.EP},
       adsurl = {https://ui.adsabs.harvard.edu/abs/2016A&A...589A..87B}
}

@ARTICLE{Gerosa+2024,
       author = {{Gerosa}, Fabiola Antonietta and {Bec}, J{\'e}r{\'e}mie and {M{\'e}heut}, H{\'e}lo{\"\i}se and {Kapoor}, Anand Utsav},
        title = "{Reduction of dust radial drift by turbulence in protoplanetary disks}",
      journal = {\aap},
         year = 2024,
        month = may,
       volume = {685},
          eid = {L4},
        pages = {L4},
          doi = {10.1051/0004-6361/202449660},
archivePrefix = {arXiv},
       eprint = {2404.11544},
 primaryClass = {astro-ph.EP},
       adsurl = {https://ui.adsabs.harvard.edu/abs/2024A&A...685L...4G}
}

@ARTICLE{GaiaCollaborationVallenari2023,
       author = {{Gaia Collaboration} and {Vallenari}, A. and {Brown}, A.~G.~A. and {Prusti}, T. and {de Bruijne}, J.~H.~J. and {Arenou}, F. and {Babusiaux}, C. and {Biermann}, M. and {Creevey}, O.~L. and {Ducourant}, C. and {Evans}, D.~W. and {Eyer}, L. and {Guerra}, R. and {Hutton}, A. and {Jordi}, C. and {Klioner}, S.~A. and {Lammers}, U.~L. and {Lindegren}, L. and {Luri}, X. and {Mignard}, F. and {Panem}, C. and {Pourbaix}, D. and {Randich}, S. and {Sartoretti}, P. and {Soubiran}, C. and {Tanga}, P. and {Walton}, N.~A. and {Bailer-Jones}, C.~A.~L. and {Bastian}, U. and {Drimmel}, R. and {Jansen}, F. and {Katz}, D. and {Lattanzi}, M.~G. and {van Leeuwen}, F. and {Bakker}, J. and {Cacciari}, C. and {Casta{\~n}eda}, J. and {De Angeli}, F. and {Fabricius}, C. and {Fouesneau}, M. and {Fr{\'e}mat}, Y. and {Galluccio}, L. and {Guerrier}, A. and {Heiter}, U. and {Masana}, E. and {Messineo}, R. and {Mowlavi}, N. and {Nicolas}, C. and {Nienartowicz}, K. and {Pailler}, F. and {Panuzzo}, P. and {Riclet}, F. and {Roux}, W. and {Seabroke}, G.~M. and {Sordo}, R. and {Th{\'e}venin}, F. and {Gracia-Abril}, G. and {Portell}, J. and {Teyssier}, D. and {Altmann}, M. and {Andrae}, R. and {Audard}, M. and {Bellas-Velidis}, I. and {Benson}, K. and {Berthier}, J. and {Blomme}, R. and {Burgess}, P.~W. and {Busonero}, D. and {Busso}, G. and {C{\'a}novas}, H. and {Carry}, B. and {Cellino}, A. and {Cheek}, N. and {Clementini}, G. and {Damerdji}, Y. and {Davidson}, M. and {de Teodoro}, P. and {Nu{\~n}ez Campos}, M. and {Delchambre}, L. and {Dell'Oro}, A. and {Esquej}, P. and {Fern{\'a}ndez-Hern{\'a}ndez}, J. and {Fraile}, E. and {Garabato}, D. and {Garc{\'\i}a-Lario}, P. and {Gosset}, E. and {Haigron}, R. and {Halbwachs}, J.-L. and {Hambly}, N.~C. and {Harrison}, D.~L. and {Hern{\'a}ndez}, J. and {Hestroffer}, D. and {Hodgkin}, S.~T. and {Holl}, B. and {Jan{\ss}en}, K. and {Jevardat de Fombelle}, G. and {Jordan}, S. and {Krone-Martins}, A. and {Lanzafame}, A.~C. and {L{\"o}ffler}, W. and {Marchal}, O. and {Marrese}, P.~M. and {Moitinho}, A. and {Muinonen}, K. and {Osborne}, P. and {Pancino}, E. and {Pauwels}, T. and {Recio-Blanco}, A. and {Reyl{\'e}}, C. and {Riello}, M. and {Rimoldini}, L. and {Roegiers}, T. and {Rybizki}, J. and {Sarro}, L.~M. and {Siopis}, C. and {Smith}, M. and {Sozzetti}, A. and {Utrilla}, E. and {van Leeuwen}, M. and {Abbas}, U. and {{\'A}brah{\'a}m}, P. and {Abreu Aramburu}, A. and {Aerts}, C. and {Aguado}, J.~J. and {Ajaj}, M. and {Aldea-Montero}, F. and {Altavilla}, G. and {{\'A}lvarez}, M.~A. and {Alves}, J. and {Anders}, F. and {Anderson}, R.~I. and {Anglada Varela}, E. and {Antoja}, T. and {Baines}, D. and {Baker}, S.~G. and {Balaguer-N{\'u}{\~n}ez}, L. and {Balbinot}, E. and {Balog}, Z. and {Barache}, C. and {Barbato}, D. and {Barros}, M. and {Barstow}, M.~A. and {Bartolom{\'e}}, S. and {Bassilana}, J.-L. and {Bauchet}, N. and {Becciani}, U. and {Bellazzini}, M. and {Berihuete}, A. and {Bernet}, M. and {Bertone}, S. and {Bianchi}, L. and {Binnenfeld}, A. and {Blanco-Cuaresma}, S. and {Blazere}, A. and {Boch}, T. and {Bombrun}, A. and {Bossini}, D. and {Bouquillon}, S. and {Bragaglia}, A. and {Bramante}, L. and {Breedt}, E. and {Bressan}, A. and {Brouillet}, N. and {Brugaletta}, E. and {Bucciarelli}, B. and {Burlacu}, A. and {Butkevich}, A.~G. and {Buzzi}, R. and {Caffau}, E. and {Cancelliere}, R. and {Cantat-Gaudin}, T. and {Carballo}, R. and {Carlucci}, T. and {Carnerero}, M.~I. and {Carrasco}, J.~M. and {Casamiquela}, L. and {Castellani}, M. and {Castro-Ginard}, A. and {Chaoul}, L. and {Charlot}, P. and {Chemin}, L. and {Chiaramida}, V. and {Chiavassa}, A. and {Chornay}, N. and {Comoretto}, G. and {Contursi}, G. and {Cooper}, W.~J. and {Cornez}, T. and {Cowell}, S. and {Crifo}, F. and {Cropper}, M. and {Crosta}, M. and {Crowley}, C. and {Dafonte}, C. and {Dapergolas}, A. and {David}, M. and {David}, P. and {de Laverny}, P. and {De Luise}, F. and {De March}, R.},
        title = "{Gaia Data Release 3. Summary of the content and survey properties}",
      journal = {\aap},
         year = 2023,
        month = jun,
       volume = {674},
          eid = {A1},
        pages = {A1},
          doi = {10.1051/0004-6361/202243940},
archivePrefix = {arXiv},
       eprint = {2208.00211},
 primaryClass = {astro-ph.GA},
       adsurl = {https://ui.adsabs.harvard.edu/abs/2023A&A...674A...1G}
}

@ARTICLE{Carrera+2021,
       author = {{Carrera}, Daniel and {Simon}, Jacob B. and {Li}, Rixin and {Kretke}, Katherine A. and {Klahr}, Hubert},
        title = "{Protoplanetary Disk Rings as Sites for Planetesimal Formation}",
      journal = {\aj},
         year = 2021,
        month = feb,
       volume = {161},
       number = {2},
          eid = {96},
        pages = {96},
          doi = {10.3847/1538-3881/abd4d9},
archivePrefix = {arXiv},
       eprint = {2008.01727},
 primaryClass = {astro-ph.EP},
       adsurl = {https://ui.adsabs.harvard.edu/abs/2021AJ....161...96C}
}

@ARTICLE{Carrera+2022,
       author = {{Carrera}, Daniel and {Thomas}, Andrew J. and {Simon}, Jacob B. and {Small}, Matthew A. and {Kretke}, Katherine A. and {Klahr}, Hubert},
        title = "{Resilience of Planetesimal Formation in Weakly Reinforced Pressure Bumps}",
      journal = {\apj},
         year = 2022,
        month = mar,
       volume = {927},
       number = {1},
          eid = {52},
        pages = {52},
          doi = {10.3847/1538-4357/ac4d28},
archivePrefix = {arXiv},
       eprint = {2108.08315},
 primaryClass = {astro-ph.EP},
       adsurl = {https://ui.adsabs.harvard.edu/abs/2022ApJ...927...52C}
}

@ARTICLE{Zhu+2019,
       author = {{Zhu}, Zhaohuan and {Zhang}, Shangjia and {Jiang}, Yan-Fei and {Kataoka}, Akimasa and {Birnstiel}, Tilman and {Dullemond}, Cornelis P. and {Andrews}, Sean M. and {Huang}, Jane and {P{\'e}rez}, Laura M. and {Carpenter}, John M. and {Bai}, Xue-Ning and {Wilner}, David J. and {Ricci}, Luca},
        title = "{One Solution to the Mass Budget Problem for Planet Formation: Optically Thick Disks with Dust Scattering}",
      journal = {\apjl},
         year = 2019,
        month = jun,
       volume = {877},
       number = {2},
          eid = {L18},
        pages = {L18},
          doi = {10.3847/2041-8213/ab1f8c},
archivePrefix = {arXiv},
       eprint = {1904.02127},
 primaryClass = {astro-ph.EP},
       adsurl = {https://ui.adsabs.harvard.edu/abs/2019ApJ...877L..18Z}
}

@dataset{Gangi+2022,
       author = {{Gangi}, M. and {Antoniucci}, S. and {Biazzo}, K. and {Frasca}, A. and {Nisini}, B. and {Alcala}, J.~M. and {Giannini}, T. and {Manara}, C.~F. and {Giunta}, A. and {Harutyunyan}, A. and {Munari}, U. and {Vitali}, F.},
        title = "{VizieR Online Data Catalog: GIARPS T Tauri spectra (GHOsT) (Gangi+, 2022)}",
 howpublished = {VizieR On-line Data Catalog: J/A+A/667/A124. Originally published in: 2022A\&A...667A.124G},
         year = 2022,
        month = oct,
          eid = {J/A+A/667/A124},
       adsurl = {https://ui.adsabs.harvard.edu/abs/2022yCat..36670124G}
}
%

\begin{appendix}
\section{Geometrical parameters}
\label{appendix:geometrical_parameters}

The geometrical parameters determine the fraction of the beam covered by the optically thick substructures. As a result, if we fix the background optical depth, different combinations of the parameters $2r/d$ and $r/z$ that yield the same projected coverage produce identical azimuthal signatures.  

After identifying a successful model through forward modelling, we used Eq.~(1) of \citet{Scardoni+2024} to compute the area covered by the rings when the disk is inclined at $i=50^\circ$, evaluated along the projected minor axis ($90^\circ$) and major axis ($0^\circ$; $A^{i=50^\circ}_{\rm rings,90^\circ}$ and $A^{i=50^\circ}_{\rm rings,0^\circ}$, respectively). Since the azimuthal modulation arises from the difference in optical-depth coverage between these two directions, we quantify the contrast through the difference in geometrical covering factors:
\begin{equation}
    \Delta A_{0^\circ-90^\circ} =  
    \frac{A^{i=50^\circ}_{\rm rings,90^\circ} - A^{i=50^\circ}_{\rm rings,0^\circ}}
    {A^{i=50^\circ}_{\rm beam}},
    \label{eq:ff}
\end{equation}
where $A^{i=50^\circ}_{\rm beam}$ is the area of the synthesised beam.

\begin{figure}
   \centering
   \includegraphics[width=1\linewidth]{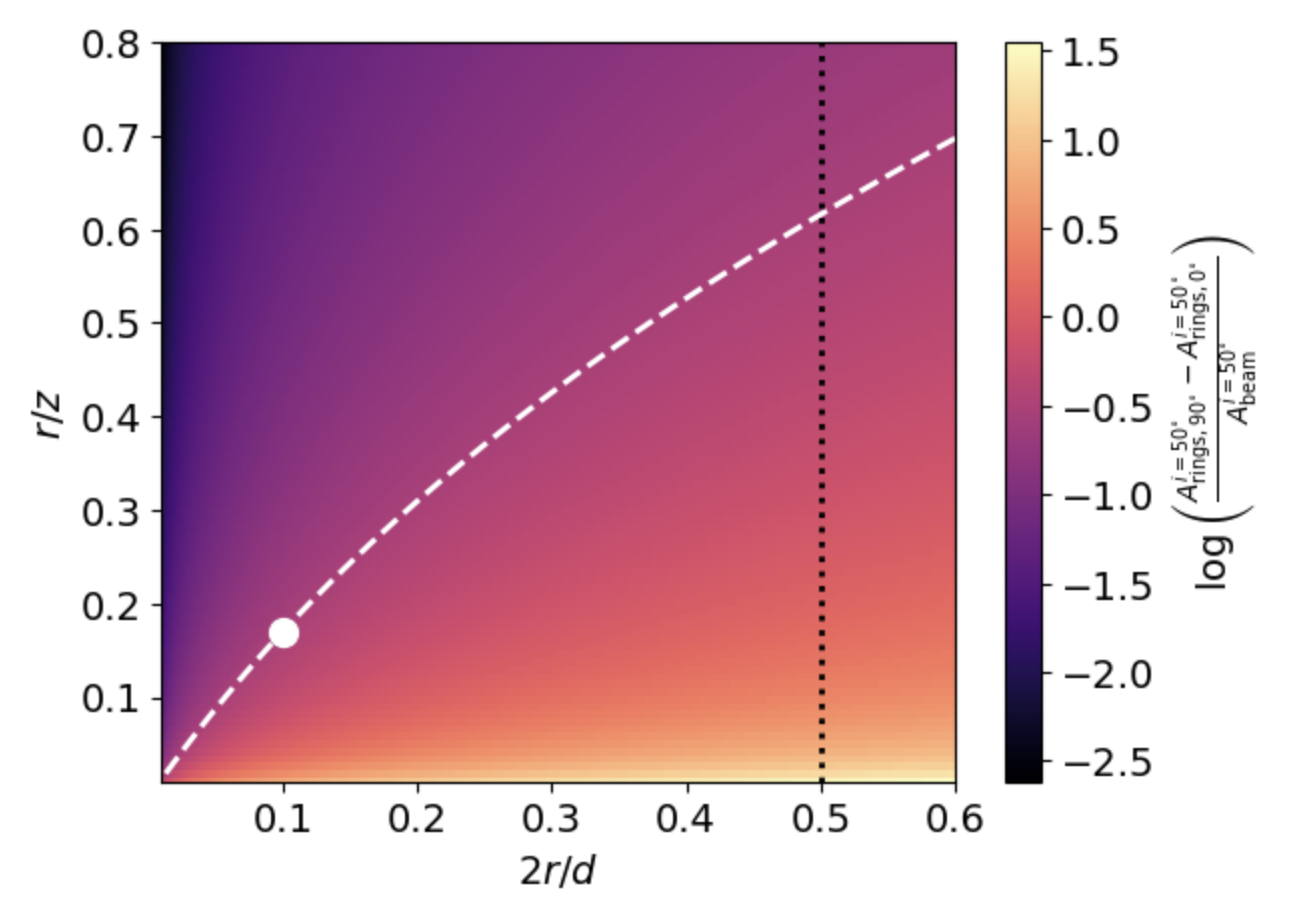}%
   \caption{Difference in filling factor, $\Delta A_{0^\circ-90^\circ}$, as a function of the geometrical parameters $2r/d$ and $r/z$. The white dotted curve indicates the combinations of parameters that reproduce the $\Delta A_{0^\circ-90^\circ}$ inferred from the successful model, marked by the white dot. The black dotted line shows an upper limit on $2r/d$, imposed by the absence of optical-depth saturation in the observed azimuthal signature.}
   \label{Fig:geomparam}
\end{figure}

We then computed $\Delta A_{0^\circ-90^\circ}$ over a grid of $2r/d$ and $r/z$ values and selected the combinations that reproduce the contrast obtained from the successful model. The results are shown in \figureautorefname~\ref{Fig:geomparam}. The white dotted curve indicates the combinations of geometrical parameters yielding the correct $\Delta A_{0^\circ-90^\circ}$, while the white dot marks the reference model. The black dotted line indicates an upper limit on $2r/d$, imposed by the absence of a flat-topped azimuthal profile (i.e., optical-depth saturation; see \citealt{Scardoni+2024}). This figure therefore illustrates the intrinsic degeneracy between the geometrical parameters $2r/d$ and $r/z$.

We caution that the geometrical parameters are constrained by the white dashed line in \figureautorefname~\ref{Fig:geomparam} only for a given value for the background optical depth ($\tau_{\nu}=0.01$ in Band 6 for this test case). Different geometrical parameters may be allowed for different background optical depth; indeed, a lower (higher) optical depth would require higher (lower) geometrical coverage of the optically thick rings.

{We can also explore geometric constraints under the assumption that the radial separation between adjacent rings cannot exceed the Band 6 beam size, i.e. $d < b_{\rm B6} \simeq 2.6~{\rm au}$ (32 mas). For the fiducial model, this implies 
$r < 0.1\,{d}/{2} < 0.13~{\rm au} \;(1.6~{\rm mas})$,
and, adopting the inferred aspect ratio, $z < {r}/{0.17} < 0.76~{\rm au} \;(9~{\rm mas})$
Considering instead the upper limit on the ratio $2r/d$, we obtain looser constraints:
$r < 0.5\,{d}/{2} < 0.65~{\rm au} \;(8~{\rm mas})$, and $z < {r}/{0.6} < 1.1~{\rm au} \;(13~{\rm mas})$.}

\section{Wavelength dependence of the azimuthal signature}
\label{appendix:Wavelength dependence of the azimuthal signature}
\begin{figure}
   \centering
   \includegraphics[width=1\linewidth]{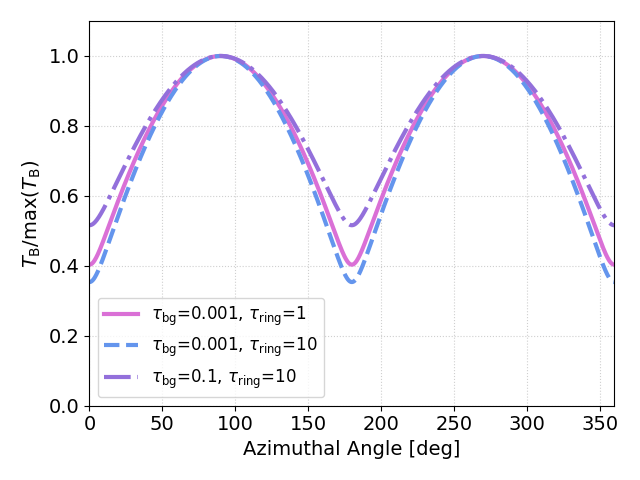}%
   \caption{{Normalised brightness temperature profiles versus azimuthal angle for three selected models. The plot illustrates the contrast variations in the curve modulation for different combinations of background ($\tau_{\rm bg}$) and ring ($\tau_{\rm ring}$) optical depths. The pink solid like is representative of Band 3, the blue dashed line is representative of Band 6, the purple dash-dotted line is representative of Band 7.}}
   \label{Fig:TauContrast}
\end{figure}
\begin{figure*}
   \centering
   \includegraphics[width=1\linewidth]{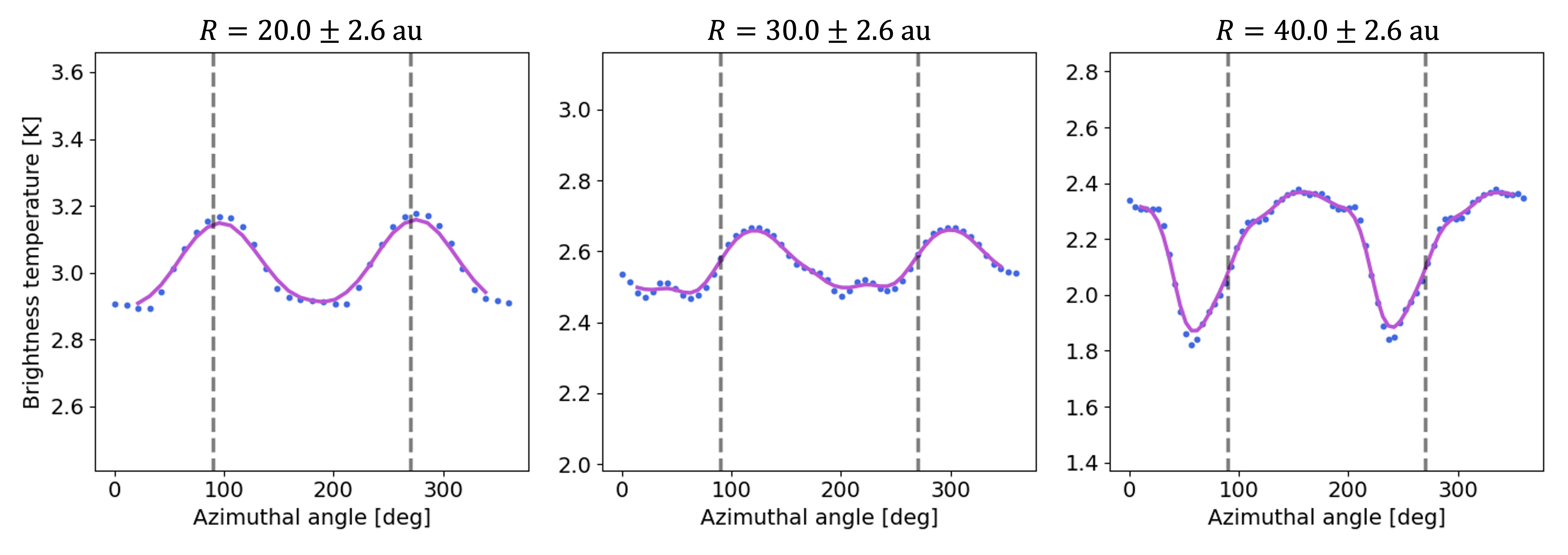}%
   \caption{Azimuthal brightness profiles of the warped disk at 20 au (left panel), 30 au (central panel), and 50 au (right panel). Each curve shows a double peaked shape, but the position of the peaks varies with radius due to the radial change in the warped disk orientation.}
   \label{Fig:Warpeddisk}
\end{figure*}

{The azimuthal modulation predicted by \citet{Scardoni+2024} arises from the contrast between optically thick unresolved rings and a more optically thin background: when the disc is observed at an inclination, the relative coverage of optically thick areas is higher at the minor axis with respect to the major axis. The signature amplitude therefore depends not only on the geometry of the unresolved rings, but also on the relative optical depths of the two components, which vary with observing wavelength through the dust opacity.} {Additionally, the different beam size among bands causes a variation in beam dilution; this artificially suppresses the emission peaks and reduces the overall contrast among bands. While this effect diminishes the observed temperature contrast between the major and minor axes compared to the intrinsic geometric models, it is not strong enough in our data to alter the expected variation in temperature modulation among the bands.}

{In general, the dust opacity scales with frequency as $\kappa_\nu \propto \nu^\beta$, implying higher optical depths at shorter wavelengths. This affects both the background and the rings. Focusing first on the rings, we expect that at relatively short wavelengths (Bands 6 and 7) they remain optically thick due to their high column density; in this regime, the resulting emission becomes insensitive to the exact value of $\tau_{\rm ring}$, justifying our adoption of a fixed representative value of 10. At longer wavelengths (Band 3), the reduced opacity leads to lower optical depths despite the high column density, so that the rings become only marginally optically thick, with $\tau_{\rm ring} \sim 5$ emerging from our modelling.}
{The background is more strongly affected by the wavelength dependence of the opacity, as it is intrinsically less dense than the rings and therefore expected to remain in the optically thin regime. Indeed, from our modelling the background optical depth increases from band 3 to 6 to 7.} 

{The resulting wavelength dependence of the azimuthal signature can be understood as follows. At Band 3, the modulation is reduced because the rings are only marginally optically thick while the background remains optically thin. At Band 6, the contrast is maximised, as the rings are optically thick and the background is still optically thin, producing the strongest modulation. At Band 7, the signal is slightly reduced due to the increased background optical depth, which lowers the ring-to-background contrast while remaining in the optically thin regime. This behavior is illustrated by the representative models in \figureautorefname~\ref{Fig:TauContrast}: the pink solid line shows a model with an optically thin background and a moderately optically thick ring (Band 3 like contrast). The blue dashed line keeps the optically thin background but increases the ring optical depth (Band 6 like contrast), thereby increasing the temperature contrast between the minor and major axis. Lastly, the purple dash-dotted line keeps the same ring optical depth as the blue line but increases the background optical depth (Band 7 like contrast), decreasing the contrast. This is consistent with the trends observed in the data and discussed in the main text.}

\section{Azimuthal brightness profiles in warped disks}
\label{appendix:warped discs}
In warped disks the local inclination and position angle of the material can change with radius; this produces a twisted appearance that can induce asymmetric azimuthal brightness. Therefore, we test here the expected azimuthal brightness radial profiles for warped disks to check whether they could reproduce the azimuthal brightness asymmetries observed in CI Tau. For this purpose, we generated a simple model of a warped disk, where the position angle varies from $60^{\circ}$ to $180^{\circ}$, while the inclination increases from $20^{\circ}$ to $60^{\circ}$ in the considered radial range (10-50 au).

We then create a synthetic observation of the warped disk following the same procedure as that in \sectionautorefname~\ref{sec:modelling the azimuthal signature} and extract the azimuthal brightness profiles at 20 au, 30 au, and 40 au (\figureautorefname~\ref{Fig:Warpeddisk}). We can notice that the azimuthal profiles of the warped disk show a double peaked structure, confirming that a warp can indeed produce an apparent brightness enhancement at the minor axis (e.g. at 20 au in the considered model). However, the radial variation in orientation due to the warp determines a shift of the position of the two peaks with radius, as we can see comparing the 3 panels in \figureautorefname~\ref{Fig:Warpeddisk}. This means that even if, in principle, a warped disk can mimic the double peaked morphology seen in observations of unresolved rings, this effect requires a precise combination of warp geometry and inclination. In contrast, emission from unresolved optically thick rings always produces a double peak along the minor axis.

Furthermore, if the double peaked structure were produced solely by geometric effects, the relative strength of the peaks should be identical across all wavelengths. This is not the case in CI Tau: the peak strength varies between Bands 3, 6, and 7. Such wavelength dependent variations are naturally explained when optical depth effects (both of the rings and the background emission) contribute to the observed azimuthal structure.

{In specific cases, the global structure of the 2D surface density map may help distinguish between scenarios, although this is not always possible and depends on both the warp properties and the observational conditions.}

Thus, although the azimuthal double peak does not uniquely indicate the presence of unresolved rings, the "fine-tuned" geometry required for warped disks makes the interpretation in terms of unresolved ring-like substructures more likely than a geometrically warped disk. 

\end{appendix}

\end{document}